\documentclass[10pt,letterpaper]{article}

\usepackage{cogsci}
\usepackage{pslatex}
\usepackage{apacite}
\usepackage{natbib}
\usepackage{amsmath,amsfonts,amssymb}
\usepackage{bm}
\usepackage{color}
\usepackage{graphicx}
\usepackage[ruled]{algorithm2e}

\usepackage{pgfplots}
\pgfplotsset{compat=newest}
\pgfplotsset{plot coordinates/math parser=false}
\newlength\fheight
\newlength\fwidth
\newlength\fheightacc
\newlength\fwidthacc
\newlength\fheightmcmc
\newlength\fwidthmcmc
\newlength\fheightcrp
\newlength\fwidthcrp
\graphicspath{{./figures/}}

\title{Probabilistic Formulation of the Take The Best Heuristic}

\author{{\large \bf Tomi Peltola (tomi.peltola@aalto.fi)} \\
  Helsinki Institute for Information Technology HIIT, Department of Computer Science\\
  Aalto University, Espoo, Finland
  \AND {\large \bf Jussi Jokinen (jussi.jokinen@aalto.fi)} \\
  Department of Communications and Networking\\
  Aalto University, Espoo, Finland
  \AND {\large \bf Samuel Kaski (samuel.kaski@aalto.fi)} \\
  Helsinki Institute for Information Technology HIIT, Department of Computer Science\\
  Aalto University, Espoo, Finland}
  
\date{December 2017}

\DeclareMathOperator{\normalpdf}{N}
\DeclareMathOperator{\bernoullipdf}{Ber}

\newcommand{\tp}{^{\mathrm{T}}}

\begin{document}

\maketitle

\begin{abstract}
The framework of cognitively bounded rationality treats problem solving as fundamentally rational, but emphasises that it is constrained by cognitive architecture and the task environment.
This paper investigates a simple decision making heuristic, Take The Best (TTB), within that framework.
We formulate TTB as a likelihood-based probabilistic model, where the decision strategy arises by probabilistic inference based on the training data and the model constraints. The strengths of the probabilistic formulation, in addition to providing a bounded rational account of the learning of the heuristic, include natural extensibility with additional cognitively plausible constraints and prior information, and the possibility to embed the heuristic as a subpart of a larger probabilistic model. We extend the model to learn cue discrimination thresholds for continuous-valued cues and experiment with using the model to account for biased preference feedback from a boundedly rational agent in a simulated interactive machine learning task.

\textbf{Keywords:} 
Bayesian models; bounded rationality; heuristics; Take The Best
\end{abstract}

\section{Introduction}

Natural environments require agents to make decisions with incomplete information and in limited time.
This, combined with the agent's limited information processing capacity, results in use of heuristic `good enough', or `satisficing', algorithms, that do not necessarily consider all possible alternatives \citep{simon1956rational}.
One such algorithm is \emph{Take The Best} (TTB), which uses subjectively ranked informative cues to discriminate between alternatives \citep{gigerenzer1996reasoning}.
It searches through the ranked list of cues, until it finds a cue that discriminates between two choices.
At this point, the search is stopped and decision made solely based on the last cue.
This type of decision-making behaviour can be analysed under the concept of \emph{bounded rationality} or \emph{computational rationality}, where agents aim to maximise expected utility of their actions, given their architectural bounds as well as those of the task environment \citep{gershman2015computational,howes2009rational,simon1956rational}.
The power of this approach is that rational behaviour arises from the structure of the environment and the task.

TTB has been shown to approximate one of the core strategies in human-level decision-making  \citep{broder2000assessing}.
\citet{schulz2016simple} showed how the heuristic decision strategy arises from its components, using an approximate Bayesian computation inspired algorithm (ABC-TTB) to fit probability distributions on the parameters. We build on this idea and formalize a \emph{Probabilistic Take The Best} model. Compared to ABC-TTB, the probabilistic TTB formulated here uses a proper likelihood-based model, with a noise model included in the model specification. With the proper likelihood model, we can use standard Bayesian computational tools for posterior inference, and the decision strategy arises from the posited model structure by conditioning on the training dataset. When used with non-informative (uniform) priors, the model has no tunable parameters. Informative prior distributions could be used to include prior knowledge.

The benefits of this approach are: (1) It lays out explicitly the assumptions in the TTB heuristic, and separates computation and model. (2) It provides a principled approach to learning the parameters of the model and their uncertainty, and suggests a possibility to extend the model without needing to change the computational framework. (3) It suggests a possibility of including prior information (or biases) in the prior distributions. (4) The TTB heuristic can be used as a component in a larger probabilistic model, such that the uncertainty of parameters in one component propagates in a natural way to other components.

Probabilistic extensions of TTB has also been considered by others. For example, \citet{lee2016bayesian} introduces a similar error model to ours, but focuses on model selection among multiple heuristics. \citet{heck2017information} considers a more elaborate error model, assuming higher rates of errors the more steps a decision takes. Neither, however, learns the cue search order and directions probabilistically. Our model could be naturally extended with more complex error models and used in model selection contexts.

In the next sections, we specify the model and give computational details. We then demonstrate the benefits of the formulation by proposing an extension for learning of cue discrimination thresholds, that is, the minimum differences in the cues for continuous or ordinal-valued cues required for discrimination. After comparing the performance to classic TTB, ABC-TTB, and logistic regression on benchmark datasets, we use the probabilistic TTB model as a component of a larger probabilistic model and, in particular, simulate a function learning case. In the case study, a boundedly rational (using TTB) agent is assumed, giving us pairwise preferences on the function evaluated at pairs of points. We show how to accommodate this bias in the function learning.

\section{Probabilistic Take The Best Model}

Let $x_i \in \mathbb{R}^M$ and $x_j\in \mathbb{R}^M$ be feature vectors of items $i$ and $j$. Let $\delta_{ij} = x_i - x_j \in \mathbb{R}^M$ be their difference. Take The Best (TTB) looks at the features, called cues, one by one in a specific order to select which item is preferred. When it finds the first cue that discriminates between the items (that is, $\delta_{ij}^{(m)} \neq 0$), it chooses one of the items as a winner depending on which direction is preferred for this cue. For example, assume the $m$th cue discriminates and $\delta_{ij}^{(m)} > 0$, and a large value is preferred for the $m$th cue. The winner is then item $i$ and the output for the comparison is $y_{ij} = 1$, whereas if a smaller value would be preferred, then $y_{ij} = 0$.

The parameters of the TTB model are the cue order and directions. These are classically learned by looking at the correlations of the cues with the criterion quantity in a training set \citep{gigerenzer1996reasoning}.

To formulate the probabilistic model, let $g$ denote the cue order (takes a value of a permutation of the $M$ cues) and $d \in \{-1,1\}^M$ the directions. For the probabilistic model to tolerate noise, we allow the outcome of the comparison to randomly flip (a flip noise likelihood), which brings in a third parameter, the flip probability $\epsilon \in (0, \frac{1}{2})$. The probabilistic model with uniform priors on $g$ and $d$ and a beta distribution prior on $\epsilon$ (restricted to $(0, \frac{1}{2})$) with parameters $1,1$ (uniform) is
\begin{align*}
  p(g) &= \frac{1}{M!}, & \\
  p(d_m) &= \frac{1}{2}, & m = 1,\ldots,M, \\
  p(\epsilon) &= 2 I(0 < \epsilon < \frac{1}{2}),& 
\end{align*}
where the indicator function $I(C) = 1$ if the condition $C$ holds and $0$ otherwise.

Let $T_{g,d}(x_i, x_j)$ give the prediction of the non-probabilistic TTB given the cue order $g$ and directions $d$. Given $g$, $d$, and $\epsilon$, the observation model (implied by the flip noise assumption) for $y_{ij}$ in the probabilistic model we introduce is
\begin{equation}
  \begin{split}
  p(y_{ij} \mid x_i, x_j, \epsilon, g, d) = &I(T_{g,d}(x_i, x_j) = 1) \bernoullipdf(y_{ij} \mid 1 - \epsilon) \\
  + &I(T_{g,d}(x_i, x_j) = 0) \bernoullipdf(y_{ij} \mid \epsilon) \\
    + &I(T_{g,d}(x_i, x_j) = \varnothing) \bernoullipdf(y_{ij} \mid \frac{1}{2}).
  \end{split} \label{eqn:obs_model}
\end{equation}
The Bernoulli ($\bernoullipdf$) distributions model the flip noise. The last branch, $T_{g,d}(x_i, x_j) = \varnothing$ corresponds to the case of no discriminating cues between $i$ and $j$, with the outcome chosen randomly with probability $\frac{1}{2}$.

The number of configurations of cue order and directions is $M! \times 2^M$. For a small number of cues, to compute the posterior distribution $p(g, d, \epsilon \mid \mathcal{D})$ given a dataset $\mathcal{D}$, one can enumerate all the possibilities. For larger numbers, Markov chain Monte Carlo sampling can be used. Computational details are given in the next section.

\section{Computational Details}

Computational details of the posterior distribution, parameter inference, and predictive distribution are given in this section. We first discuss how to simplify the likelihood and integrate out the flip noise parameter $\epsilon$ analytically, allowing posterior computation to focus on the cue search order and directions. We give a Markov chain Monte Carlo algorithm for parameter inference for cases where exhaustive computation over the cue orders and directions is infeasible.

\subsection{Marginalizing $\epsilon$ and Posterior Distribution}

The likelihood given $g$, $d$, and $\epsilon$ is given by Equation~\ref{eqn:obs_model}. This simplifies to $\epsilon$ for the pairs $(y_{ij} = 0, T_{g,d}(x_i, x_j) = 1)$ and $(y_{ij} = 1, T_{g,d}(x_i, x_j) = 0)$, that is, when the observed value and predicted value are different (called ``wrong predictions'' below), and to $1 - \epsilon$ for $(y_{ij} = 1, T_{g,d}(x_i, x_j) = 1)$ and $(y_{ij} = 0, T_{g,d}(x_i, x_j) = 0)$, that is, when observed and predicted value are the same (called ``correct predictions'' below). When $T_{g,d}(x_i, x_j) = \varnothing$ (called ``undecided predictions'' below), the likelihood of either outcome is $\frac{1}{2}$.

We can then compute the likelihood for multiple observations of comparisons from a set of pairs $\mathcal{P}$, $Y = \{y_{ij}; (i,j) \in \mathcal{P}\}$, $X = \{(x_i, x_j); (i,j) \in \mathcal{P}\}$, and marginalize out $\epsilon$:
\begin{equation}
  \begin{split}
    p(Y \mid X, g, d) &= \int_0^{\frac{1}{2}} \prod_{(i,j) \in \mathcal{P}} p(y_{ij} \mid x_i, x_j, \epsilon, g, d) p(\epsilon) d\epsilon \\
    &= \frac{1}{Z_\epsilon} \left(\frac{1}{2}\right)^{N_\varnothing} \int_0^{\frac{1}{2}} \epsilon^{N_i + \alpha - 1} (1 - \epsilon)^{N_c + \beta - 1} d\epsilon \\
    &= \frac{1}{Z_\epsilon} \left(\frac{1}{2}\right)^{N_\varnothing} B_{\frac{1}{2}}(N_i + \alpha, N_c + \beta),
  \end{split} \label{eqn:likelihood}
\end{equation}
where $N_\varnothing$, $N_i$, and $N_c$ are the numbers of undecided, incorrect, and correct predictions by the TTB model with the cue order $g$ and directions $d$. Here, $Z_\epsilon$ is the normalizing constant of the beta prior with parameters $\alpha$ and $\beta$ restricted to $(0, \frac{1}{2})$ (where $\alpha = 1$, $\beta = 1$ for the uniform prior), and $B_{\frac{1}{2}}$ is the incomplete beta function\footnote{``Unregularized'' incomplete beta function, as defined by the integral in the second-to-last line of Equation~\ref{eqn:likelihood}.}.

Let $\mathcal{D} = (Y, X)$ be the observed training set. The posterior probabilities $p(g, d \mid \mathcal{D})$ are proportional to Equation~\ref{eqn:likelihood}, with the sum $Z = \sum_{g,d} p(g, d \mid \mathcal{D})$ being the normalizing constant.

The conditional posterior distribution of $\epsilon$ given $g, d$ is
\begin{equation*}
  p(\epsilon \mid \mathcal{D}, g, d) = \frac{\epsilon^{N_i + \alpha - 1} (1 - \epsilon)^{N_c + \beta - 1}}{B_{\frac{1}{2}}(N_i + \alpha, N_c + \beta)}.
\end{equation*}
This is a beta distribution with parameters $N_i + \alpha$ and $N_c + \beta$ restricted to $(0, \frac{1}{2})$.

\subsection{Markov Chain Monte Carlo Sampling}

A collapsed Gibbs sampling algorithm is used. We first integrate over the noise flip probability $\epsilon$ and only sample over the cue order $g$ and directions $d$ using Algorithm~\ref{alg:mcmc}.

\begin{algorithm}[h]
\SetAlgoLined
\KwData{$\mathcal{D} = (Y, X)$ and $g$ and $d$ priors.}
\KwResult{$S$ samples from $p(g, d \mid \mathcal{D})$.}
 Initialize $g$ and $d$, e.g., randomly.\;
 \For{$s \gets 1$ \KwTo $S$}{
  \For{$m \gets 1$ \KwTo $M$ in random order}{
  1. Form all cue orders such that the cue $m$ takes all positions and the other cues remain in the same order, giving $M$ different cue orders.\;
  2. With the cue orders above, form all TTB models with those orders and with the cue $m$ taking either direction and the directions of the other cues kept constant, giving $2 M$ models (configurations of $g$ and $d$).\;
  3. Sample the next configuration from these with probabilities proportional to their unnormalized posterior probabilities, that is, $\propto p(g, d \mid \mathcal{D})$.\;
  }
  Save the configuration of $g$ and $d$ as the $s$th sample.\;
 }
 \caption{Markov chain Monte Carlo sampler.}\label{alg:mcmc}
\end{algorithm}

\subsection{Predictive Distribution}

Given a pair of new data points $\tilde{x}_1$ and $\tilde{x}_2$, the posterior predictive distribution of their comparison $\tilde{y}$ is
\begin{equation*}
  \begin{split}
    & p(\tilde{y} \mid \tilde{x}_1, \tilde{x}_2, \mathcal{D})\\ &= \sum_{g,d} \int_0^{\frac{1}{2}} p(\tilde{y} \mid \tilde{x}_1, \tilde{x}_2, \epsilon, g, d) p(\epsilon, g, d \mid \mathcal{D}) d\epsilon \\
    &= \sum_{g,d} p(g, d \mid \mathcal{D}) \int_0^{\frac{1}{2}} p(\tilde{y} \mid \tilde{x}_1, \tilde{x}_2, \epsilon, g, d) p(\epsilon \mid g, d, \mathcal{D}) d\epsilon \\
    &\approx \frac{1}{S} \sum_{s=1}^S \int_0^{\frac{1}{2}} p(\tilde{y} \mid \tilde{x}_1, \tilde{x}_2, \epsilon, g^{(s)}, d^{(s)}) p(\epsilon \mid g^{(s)}, d^{(s)}, \mathcal{D}) d\epsilon,
  \end{split}
\end{equation*}
where the last line applies in the case we have sampled $S$ samples $(g^{(s)}, d^{(s)})$ from the posterior $p(g, d \mid \mathcal{D})$.

The integral over $\epsilon$ again gives incomplete beta functions
\begin{equation*}
\begin{split}
  &\int_0^{\frac{1}{2}} p(\tilde{y} \mid \tilde{x}_1, \tilde{x}_2, \epsilon, g, d) p(\epsilon \mid g, d, \mathcal{D}) d\epsilon\\ & = \left(\frac{1}{2}\right)^{I_\varnothing} \frac{B_{\frac{1}{2}}(N_i + \alpha + I_i, N_c + \beta + I_c)}{B_{\frac{1}{2}}(N_i + \alpha, N_c + \beta)},
  \end{split}
\end{equation*}
where $I_\varnothing = I(T_{g,d}(\tilde{x}_1, \tilde{x}_2) = \varnothing)$, $I_i = I(T_{g,d}(\tilde{x}_1, \tilde{x}_2) \neq \tilde{y})$, and $I_c = I(T_{g,d}(\tilde{x}_1, \tilde{x}_2) = \tilde{y})$.

\section{Extensions}

\subsection{Handling Correlation in the Comparisons}

In the above, we have assumed that the $y_{ij}$ are independent (conditional on the model parameters). This, however, would often be expected to be violated. For example, we might often assume (some degree of) transitivity: if $i$ is preferred to $j$ and $j$ to $k$, $i$ could be assumed to be preferred to $k$.

In this work, we use a heuristic approach to account for transitivity dependencies by downweighting the likelihood terms of each $y_{ij}$. When we observe the full set of pairwise comparisons for $N$ items, we assume that the ranking of the items would be enough to decide all pairwise comparisons. Since each pairwise comparison is 1 bit of information and there are $N!$ ways to rank $N$ items (leading to $\log_2 N!$ bits of information), we assign a weight $\frac{\log_2 N!}{{N \choose 2}}$ for each likelihood term (the terms are raised to this power). More formal ways of dealing with dependencies are left for future work.

\subsection{Cue Discrimination Thresholds}

For real-valued cues, practically all elements of $\delta_{ij} = x_i - x_j$ will be non-zero and the first cue will always discriminate. Yet, a small difference in a cue might be non-informative or imperceptible for humans (especially, if the cues are not given as precise numbers but, for example, visually), and large differences are more salient. We can extend the model to include non-negative threshold parameters $t_m \in [0, \infty)$ for each cue, such that only differences $|\delta_{ij}^{(m)}| > t_m$ are considered to discriminate between the items.

We have extended the MCMC sampling to allow sampling over a discrete set of pre-specified thresholds for each cue (assumed to have uniform prior). Extension to continuous-valued thresholds would also be possible.

\section{Results}

We first establish that the probabilistic Take The Best model is effective at learning from training data, and then demonstrate the model as a part of a larger probabilistic model.

\subsection{Performance on Benchmark Datasets}

\begin{figure*}[!htb]
    \centering
\input{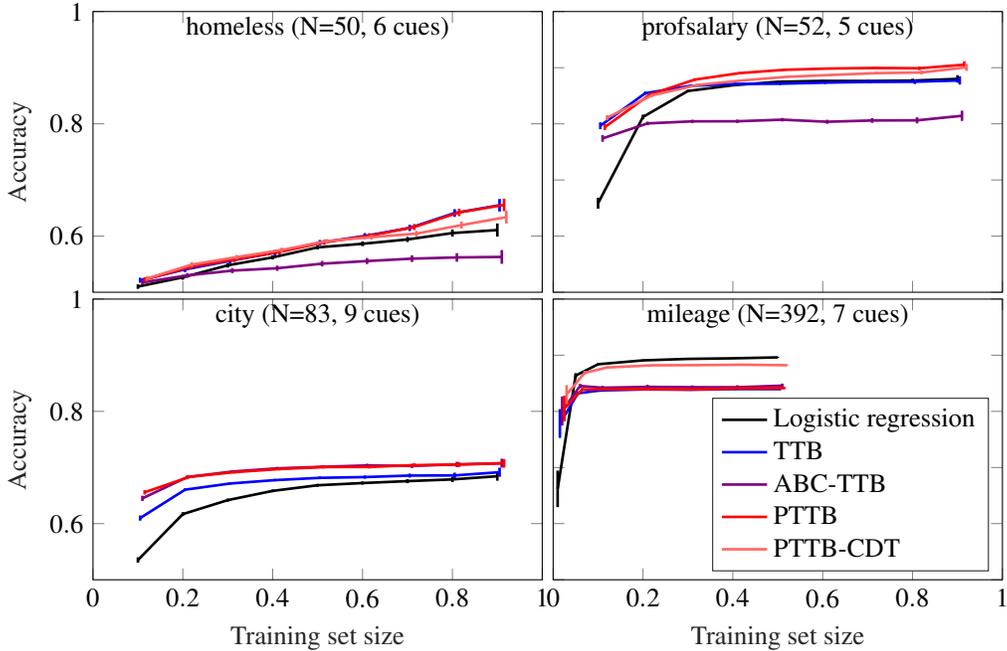}
\vspace{-5mm}\caption{Accuracy on homeless, profsalary, city, and mileage datasets as a function of training dataset size (fraction of full data). Mileage data has a large number of samples and was run only up to 50\% of full data (with fractions 0.01, 0.05 included). TTB=Take The Best; ABC-TTB=Approximately Bayesian Computed Take The Best \citep{schulz2016simple}; PTTB=Probabilistic TTB; PTTB-CDT=PTTB with cue discrimination threshold learning.}
    \label{fig:acc}
\end{figure*}

The accuracy of the probabilistic TTB (PTTB) model is compared with the classic TTB, ABC-TTB, and logistic regression on four benchmark datasets. Heuristica R-package\footnote{\url{https://cran.r-project.org/web/packages/heuristica/}, version 1.0.1.} is used for the classic TTB and logistic regression. ABC-TTB code is from \url{https://github.com/ericschulz/TTBABC}\footnote{With minor bug fixes.}. We used the same parameters
for ABC-TTB as were used for the city dataset by \citet{schulz2016simple} and made no effort to tune them for the other datasets. The other models have no tuning parameters. MCMC computation with 1000 samples after burn-in of 100 was used for PTTB models. The datasets homeless, profsalary, and city were obtained from \url{https://github.com/ericschulz/TTBABC} and mileage dataset is from Matlab (``carbig.mat'' with data points containing missing data values removed).

Figure~\ref{fig:acc} shows the discrimination accuracies as a function of training set size over 1000 replications of training and test set splits. The PTTB performs comparably or better compared to TTB and ABC-TTB methods. It also outperforms logistic regression except in the mileage dataset, where logistic regression is the best method expect in the smallest tested training set size case. The cue discrimination threshold learning (PTTB-CDT) decreases performance modestly in homeless and profsalary datasets, but increases performance considerably in the mileage dataset.

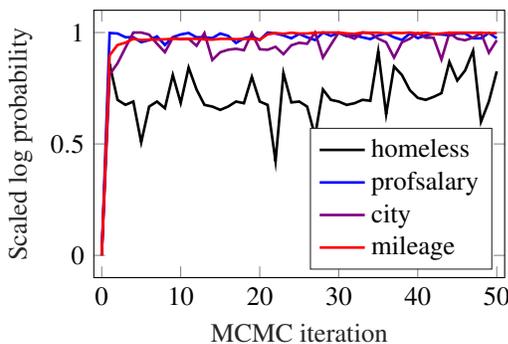
\begin{figure}[!htb]
    \centering
%
%
\begin{tikzpicture}

\begin{axis}[%
width=0.96\fwidthmcmc,
height=\fheightmcmc,
at={(0\fwidthmcmc,0\fheightmcmc)},
scale only axis,
xmin=-1,
xmax=51,
xlabel style={font=\color{white!15!black}},
xlabel={MCMC iteration},
ymin=-0.1,
ymax=1.1,
ylabel style={font=\color{white!15!black}},
ylabel={Scaled log probability},
axis background/.style={fill=white},
legend style={at={(0.97,0.03)}, anchor=south east, legend cell align=left, align=left, draw=white!15!black}
]
\addplot [color=black, line width=1pt]
  table[row sep=crcr]{%
0	0\\
1	0.858331307582358\\
2	0.697879450579091\\
3	0.675432382341529\\
4	0.690272167441278\\
5	0.507305363611281\\
6	0.667723133207649\\
7	0.690272167441278\\
8	0.660297571081892\\
9	0.809185981816622\\
10	0.683862063816501\\
11	0.8427705850901\\
12	0.744481326818907\\
13	0.67519405561776\\
14	0.667723133207649\\
15	0.652917323013006\\
16	0.667723133207649\\
17	0.690272167441278\\
18	0.682710384785785\\
19	0.800936590372352\\
20	0.690272167441278\\
21	0.67519405561776\\
22	0.422806805497916\\
23	0.817481615573228\\
24	0.68634977846373\\
25	0.690272167441278\\
26	0.667723133207649\\
27	0.535240138756217\\
28	0.744481326818907\\
29	0.697879450579091\\
30	0.690272167441278\\
31	0.67519405561776\\
32	0.682710384785785\\
33	0.697879450579091\\
34	0.693777898291199\\
35	0.915420021539273\\
36	0.633177284346137\\
37	0.847598959949547\\
38	0.809185981816622\\
39	0.741160214737379\\
40	0.706723317096864\\
41	0.697879450579091\\
42	0.711173790227542\\
43	0.727932679912196\\
44	0.868229455138522\\
45	0.784576327494983\\
46	0.830037198219613\\
47	0.917578343656885\\
48	0.599508844746374\\
49	0.690272167441278\\
50	0.82582354382837\\
};
\addlegendentry{homeless}

\addplot [color=blue, line width=1pt]
  table[row sep=crcr]{%
0	0\\
1	0.997940226563204\\
2	0.995672549564909\\
3	0.979054849868335\\
4	0.978529955451275\\
5	0.956733468663642\\
6	0.965845359139707\\
7	0.98279075636193\\
8	0.944248607971468\\
9	0.979054849868335\\
10	0.991361937585268\\
11	0.997940226563204\\
12	0.978529955451275\\
13	0.978529955451275\\
14	0.974829978769475\\
15	0.995672549564909\\
16	0.985048802371303\\
17	0.953305186634259\\
18	0.974009706505654\\
19	0.98279075636193\\
20	0.966428235832365\\
21	1\\
22	0.995672549564909\\
23	0.987151772012231\\
24	0.974009706505654\\
25	0.993626406938645\\
26	0.978529955451275\\
27	0.966428235832365\\
28	1\\
29	0.985048802371303\\
30	0.997940226563204\\
31	0.983295864192086\\
32	0.987068045434294\\
33	0.978529955451275\\
34	0.970621142883411\\
35	0.987068045434294\\
36	0.995672549564909\\
37	0.978529955451275\\
38	0.966428235832365\\
39	0.995672549564909\\
40	0.974829978769475\\
41	0.987068045434294\\
42	0.993626406938645\\
43	0.991361937585268\\
44	0.987553131424527\\
45	0.995672549564909\\
46	0.995672549564909\\
47	0.974285529576332\\
48	0.974009706505654\\
49	0.997940226563204\\
50	0.974285529576332\\
};
\addlegendentry{profsalary}

\addplot [color=violet, line width=1pt]
  table[row sep=crcr]{%
0	0\\
1	0.811620503687846\\
2	0.861808225341039\\
3	0.931567868146337\\
4	1\\
5	1\\
6	0.990702868073465\\
7	0.94952570756719\\
8	0.911614768802329\\
9	0.976857571576298\\
10	0.969979901907705\\
11	0.972269135839644\\
12	0.940520587957036\\
13	0.997670681257319\\
14	0.876787880101932\\
15	0.911614768802329\\
16	0.922667341427742\\
17	0.927111093469199\\
18	0.920450340939689\\
19	0.997670681257319\\
20	0.924887591212622\\
21	0.922667341427742\\
22	0.936037691056486\\
23	0.902830933968646\\
24	0.94276694724699\\
25	0.954047982603805\\
26	0.94952570756719\\
27	1\\
28	0.887583120125678\\
29	0.945016584766769\\
30	1\\
31	0.986074410173279\\
32	0.972269135839644\\
33	1\\
34	0.997670681257319\\
35	0.972269135839644\\
36	0.956314060457599\\
37	0.887583120125678\\
38	0.956314060457599\\
39	0.972269135839644\\
40	0.997670681257319\\
41	0.976857571576298\\
42	0.976857571576298\\
43	0.878940546495192\\
44	0.997670681257319\\
45	0.972269135839644\\
46	0.965411384960932\\
47	0.976857571576298\\
48	0.997670681257319\\
49	0.90941396751872\\
50	0.965411384960932\\
};
\addlegendentry{city}

\addplot [color=red, line width=1pt]
  table[row sep=crcr]{%
0	0\\
1	0.897658026775396\\
2	0.944136472718954\\
3	0.954477590839267\\
4	0.971615347148798\\
5	0.966170592911416\\
6	0.968096644269911\\
7	0.968096644269911\\
8	0.969787071042712\\
9	0.971615347148798\\
10	0.971615347148798\\
11	0.971615347148798\\
12	0.971615347148798\\
13	0.968965297897574\\
14	0.968096644269911\\
15	0.971615347148798\\
16	0.971615347148798\\
17	0.970109097463367\\
18	0.971615347148798\\
19	0.971615347148798\\
20	0.968096644269911\\
21	0.991934041833661\\
22	0.997643707794184\\
23	0.994237527993603\\
24	0.999057067795498\\
25	0.995441840738691\\
26	0.995791695789884\\
27	0.998841676443586\\
28	1\\
29	1\\
30	1\\
31	0.996062234304765\\
32	0.994257975353043\\
33	1\\
34	0.998971387044298\\
35	0.999142755269642\\
36	0.995293460051956\\
37	0.9989272959103\\
38	0.99758203080486\\
39	1\\
40	0.995625229838362\\
41	0.999228449467754\\
42	0.993569596892156\\
43	0.99570497841706\\
44	0.998971387044298\\
45	0.999057067795498\\
46	0.999142755269642\\
47	0.999057067795498\\
48	0.998885713015003\\
49	0.99733196613767\\
50	0.996753867525275\\
};
\addlegendentry{mileage}

\end{axis}
\end{tikzpicture}%
\vspace{-5mm}\caption{Scaled logarithm of the (unnormalized) posterior probability for 50 first iterations of MCMC in each dataset, starting from a random initial setting of parameters. For clear visualization in one figure, each curve is scaled (over full trace) such that its lowest value is 0 and highest is 1.}
    \label{fig:mcmctraces}
\end{figure}

\begin{figure}[!htb]
    \centering
\input{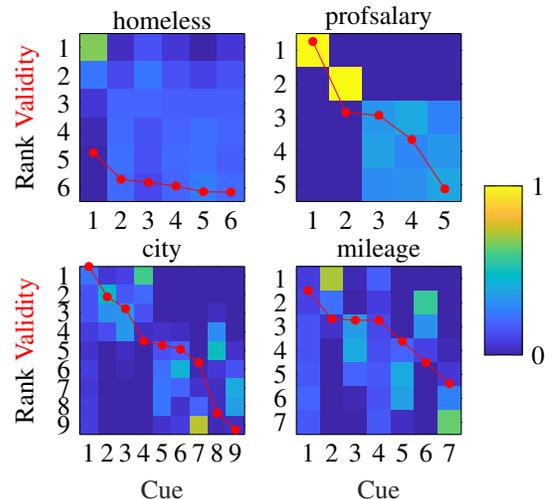}
\vspace{-5mm}\caption{PTTB cue rank posterior probabilities (heatmap) and TTB cue validities (red line; y-axis runs from 0.5 to 1) learned from full datasets.}
    \label{fig:crp}
\end{figure}

To examine the computational burden of finding good configurations of cue order and directions, Figure~\ref{fig:mcmctraces} shows the logarithm of the posterior probability of the first 50 iterations of a single chain of MCMC for each of the datasets, starting from a random configuration. Already a single iteration of the MCMC algorithm is enough to locate good models. Figure~\ref{fig:crp} shows the posterior uncertainty in the cue search order for PTTB, comparing it to the TTB cue validities. Although they have roughly similar trends, there are also clear differences.

\subsection{TTB as a Part of a Larger Probabilistic Model: Linear Regression with Pairwise Observations}

We simulate a situation where we are interested in learning a function and there is an agent which can give us preferences over pairwise evaluations of the function, but the preferences are generated through a biased mechanism, perhaps due to the limited cognitive abilities of the agent. More specifically, we consider learning a linear regression function $f(x) = w\tp x$, with regression weights $w$, where we have a few direct $(y_i, x_i)$ observations of the regression and a set of pairwise observations of whether $f(x_i)$ is larger or smaller than $f(x_j)$, generated using the TTB heuristic.

The main task is to learn the posterior distribution of $w$, $p(w \mid \mathcal{D}, \mathcal{H})$, where $\mathcal{D} = \{(y_i, x_i); i=1,\ldots,N\}$ and $\mathcal{H} = \{h_{ij}=(f^*(x_i) > f^*(x_j)); (i,j) \in \mathcal{P}\}$, where $\mathcal{P}$ is a set of pairs, where pairwise preference observations are available, and the superscript $^*$ reminds us that the observations are biased. We use a Gaussian linear regression with a Gaussian prior on $w$:
\vspace{-1mm}\begin{align*}
  p(y_i \mid x_i, w) &= \normalpdf(y_i \mid w\tp x_i, \sigma^2), & i=1,\ldots,N, \\
   p(w) &= \normalpdf(w \mid 0, \tau^2), &
\end{align*}
where we fix $\sigma^2 = 1$ and $\tau^2 = 1$ and generate a simulated dataset $\mathcal{D}$ by sampling from the true model. The posterior distribution given $\mathcal{D}$ is $p(w \mid \mathcal{D}) \propto p(w) \prod_i p(y_i \mid x_i, w)$ (which is a multivariate Gaussian distribution).

To simulate an agent generating the pairwise observations, we first form a grid of points in the covariate space, denoting the set of points $X_G$. We then form all pairwise comparisons between these points given the true function $f$ (or equivalently the true weights $w$) and use them as a training set to learn a TTB model. The observations $\mathcal{H}$ are then predictions from this TTB at a subset of the grid points. This simulates an agent who knows the true function, but can only access it through the heuristic TTB model. (We further included a threshold such that two covariate values must be further apart than it for them to discriminate in the TTB model.)

To use the pairwise observations to learn about $w$, we extend the model to include $\mathcal{H}$ through the probabilistic TTB model, similar to what was used to generate the observations: 
\begin{equation*}
\prod_{(i,j) \in \mathcal{P}} p(h_{ij} \mid x_i, x_j, \epsilon, g, d) p(\epsilon, g, d \mid X_G, w),
\end{equation*}
where the latter term is the posterior of a TTB model, where all pairwise comparisons generated by using $w$ to predict the function values in the grid $X_G$ are used as the training data. We further marginalize the TTB parameters to get the factor $p(\mathcal{H} \mid X_{\mathcal{H}}, X_G, w)$, where $X_{\mathcal{H}}$ denotes the set of pairs of points for which we have pairwise observations. That is, the posterior of $w$ is now
\begin{equation*}
  \begin{split}
    p(w \mid \mathcal{D}, \mathcal{H}) &\propto p(w) \prod_i p(y_i \mid x_i, w) p(\mathcal{H} \mid X_{\mathcal{H}}, X_G, w)\\
    &= p(w) \prod_i p(y_i \mid x_i, w) \times \\
    &\sum_{g,d} \int_0^{\frac{1}{2}}\!\!\!\!\! \prod_{(i,j) \in \mathcal{P}}\!\!p(h_{ij} \mid x_i, x_j, \epsilon, g, d) p(\epsilon, g, d \mid X_G, w) d\epsilon.
  \end{split}
\end{equation*}
In other words, the likelihood term $p(\mathcal{H} \mid X_{\mathcal{H}}, X_G, w)$ models how well the observed pairwise preferences $\mathcal{H}$ are concordant with a probabilistic TTB model induced by $w$ on the points $X_G$. The regions with values of $w$ with high concordance will get higher posterior mass. Exact computation of the probabilistic TTB model by enumeration is used in this experiment.

For comparison, we also formulate a model that includes the pairwise observations, but assumes them unbiased. For this we use the flip-noise likelihood
\begin{equation*}
  \begin{split}
    p(h_{ij} \mid x_i, x_j, w, \kappa) &= I((x_i - x_j)\tp w > 0) \bernoullipdf(h_{ij} \mid 1 - \kappa) \\
    &+ I((x_i - x_j)\tp w < 0) \bernoullipdf(h_{ij} \mid \kappa) \\
    &+ I((x_i - x_j)\tp w = 0) \bernoullipdf(h_{ij} \mid \frac{1}{2})
  \end{split}
\end{equation*}
with uniform prior on the flip-noise parameter $\kappa \in (0, \frac{1}{2})$. The product of these terms over the observations in $\mathcal{H}$ is marginalized over $\kappa$ to get the factor $p(\mathcal{H} \mid X_{\mathcal{H}}, w)$ for this alternative model.

\begin{figure}[!htb]
\hspace*{-5mm}
%
%
\begin{tikzpicture}

\begin{axis}[%
width=0.326\fwidth,
height=0.494\fheight,
at={(-0\fwidth,0.506\fheight)},
scale only axis,
axis on top,
xmin=-2.00869565217391,
xmax=2.00869565217391,
xlabel style={font=\color{white!15!black}},
xlabel={$w_2$},
ymin=-2.00869565217391,
ymax=2.00869565217391,
ylabel style={font=\color{white!15!black},yshift=-8pt},
ylabel={$w_1$},
axis background/.style={fill=white},
legend style={legend cell align=left, align=left, draw=white!15!black}
]
\addplot [forget plot] graphics [xmin=-2.00869565217391, xmax=2.00869565217391, ymin=-2.00869565217391, ymax=2.00869565217391] {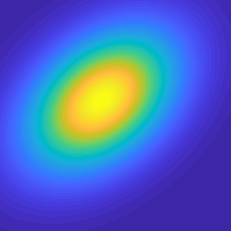};
\node[align=center, font=\color{white}]
at (axis cs:0,-1.8) {No pairwise obs.};
\addplot [color=red, draw=none, mark=o, mark options={solid, red}]
  table[row sep=crcr]{%
0.8	1\\
};

\end{axis}

\begin{axis}[%
width=0.225\fwidth,
height=0.494\fheight,
at={(0\fwidth,0\fheight)},
scale only axis,
point meta min=0,
point meta max=1,
xmin=0,
xmax=1,
xtick={0,0.5,1},
xticklabels={\empty},
ymin=0,
ymax=1,
ytick={0,0.2,0.4,0.6,0.8,1},
yticklabels={\empty},
axis line style={draw=none},
ticks=none,
axis x line*=bottom,
axis y line*=left,
legend style={legend cell align=left, align=left, draw=white!15!black},
colormap={mymap}{[1pt] rgb(0pt)=(0.2422,0.1504,0.6603); rgb(1pt)=(0.25039,0.164995,0.707614); rgb(2pt)=(0.257771,0.181781,0.751138); rgb(3pt)=(0.264729,0.197757,0.795214); rgb(4pt)=(0.270648,0.214676,0.836371); rgb(5pt)=(0.275114,0.234238,0.870986); rgb(6pt)=(0.2783,0.255871,0.899071); rgb(7pt)=(0.280333,0.278233,0.9221); rgb(8pt)=(0.281338,0.300595,0.941376); rgb(9pt)=(0.281014,0.322757,0.957886); rgb(10pt)=(0.279467,0.344671,0.971676); rgb(11pt)=(0.275971,0.366681,0.982905); rgb(12pt)=(0.269914,0.3892,0.9906); rgb(13pt)=(0.260243,0.412329,0.995157); rgb(14pt)=(0.244033,0.435833,0.998833); rgb(15pt)=(0.220643,0.460257,0.997286); rgb(16pt)=(0.196333,0.484719,0.989152); rgb(17pt)=(0.183405,0.507371,0.979795); rgb(18pt)=(0.178643,0.528857,0.968157); rgb(19pt)=(0.176438,0.549905,0.952019); rgb(20pt)=(0.168743,0.570262,0.935871); rgb(21pt)=(0.154,0.5902,0.9218); rgb(22pt)=(0.146029,0.609119,0.907857); rgb(23pt)=(0.138024,0.627629,0.89729); rgb(24pt)=(0.124814,0.645929,0.888343); rgb(25pt)=(0.111252,0.6635,0.876314); rgb(26pt)=(0.0952095,0.679829,0.859781); rgb(27pt)=(0.0688714,0.694771,0.839357); rgb(28pt)=(0.0296667,0.708167,0.816333); rgb(29pt)=(0.00357143,0.720267,0.7917); rgb(30pt)=(0.00665714,0.731214,0.766014); rgb(31pt)=(0.0433286,0.741095,0.73941); rgb(32pt)=(0.0963952,0.75,0.712038); rgb(33pt)=(0.140771,0.7584,0.684157); rgb(34pt)=(0.1717,0.766962,0.655443); rgb(35pt)=(0.193767,0.775767,0.6251); rgb(36pt)=(0.216086,0.7843,0.5923); rgb(37pt)=(0.246957,0.791795,0.556743); rgb(38pt)=(0.290614,0.79729,0.518829); rgb(39pt)=(0.340643,0.8008,0.478857); rgb(40pt)=(0.3909,0.802871,0.435448); rgb(41pt)=(0.445629,0.802419,0.390919); rgb(42pt)=(0.5044,0.7993,0.348); rgb(43pt)=(0.561562,0.794233,0.304481); rgb(44pt)=(0.617395,0.787619,0.261238); rgb(45pt)=(0.671986,0.779271,0.2227); rgb(46pt)=(0.7242,0.769843,0.191029); rgb(47pt)=(0.773833,0.759805,0.16461); rgb(48pt)=(0.820314,0.749814,0.153529); rgb(49pt)=(0.863433,0.7406,0.159633); rgb(50pt)=(0.903543,0.733029,0.177414); rgb(51pt)=(0.939257,0.728786,0.209957); rgb(52pt)=(0.972757,0.729771,0.239443); rgb(53pt)=(0.995648,0.743371,0.237148); rgb(54pt)=(0.996986,0.765857,0.219943); rgb(55pt)=(0.995205,0.789252,0.202762); rgb(56pt)=(0.9892,0.813567,0.188533); rgb(57pt)=(0.978629,0.838629,0.176557); rgb(58pt)=(0.967648,0.8639,0.16429); rgb(59pt)=(0.96101,0.889019,0.153676); rgb(60pt)=(0.959671,0.913457,0.142257); rgb(61pt)=(0.962795,0.937338,0.12651); rgb(62pt)=(0.969114,0.960629,0.106362); rgb(63pt)=(0.9769,0.9839,0.0805)},
colorbar horizontal,
colorbar style={xtick={0,1}, xticklabels={lower, higher},
                at={(0.1,0.6)},anchor=north west,
                width=64pt}
]
\end{axis}

\begin{axis}[%
width=0.326\fwidth,
height=0.494\fheight,
at={(0.337\fwidth,0.506\fheight)},
scale only axis,
axis on top,
xmin=-2.00869565217391,
xmax=2.00869565217391,
xtick={-2,-1,0,1,2},
xticklabels={\empty},
ymin=-2.00869565217391,
ymax=2.00869565217391,
ytick={-2,-1.5,-1,-0.5,0,0.5,1,1.5,2},
yticklabels={\empty},
axis background/.style={fill=white},
legend style={legend cell align=left, align=left, draw=white!15!black}
]
\addplot [forget plot] graphics [xmin=-2.00869565217391, xmax=2.00869565217391, ymin=-2.00869565217391, ymax=2.00869565217391] {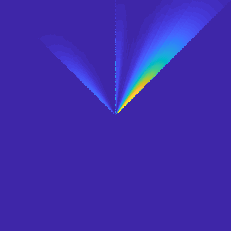};
\node[align=center, font=\color{white}]
at (axis cs:0,-1.4) {TTB obs.\\ \& TTB model};
\addplot [color=red, draw=none, mark=o, mark options={solid, red}]
  table[row sep=crcr]{%
0.8	1\\
};

\end{axis}

\begin{axis}[%
width=0.326\fwidth,
height=0.494\fheight,
at={(0.674\fwidth,0.506\fheight)},
scale only axis,
axis on top,
xmin=-2.00869565217391,
xmax=2.00869565217391,
xtick={-2,-1,0,1,2},
xticklabels={\empty},
ymin=-2.00869565217391,
ymax=2.00869565217391,
ytick={-2,-1.5,-1,-0.5,0,0.5,1,1.5,2},
yticklabels={\empty},
axis background/.style={fill=white},
legend style={legend cell align=left, align=left, draw=white!15!black}
]
\addplot [forget plot] graphics [xmin=-2.00869565217391, xmax=2.00869565217391, ymin=-2.00869565217391, ymax=2.00869565217391] {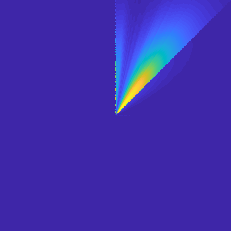};
\node[align=center, font=\color{white}]
at (axis cs:0,-1.4) {Unbiased obs.\\ \& TTB model};
\addplot [color=red, draw=none, mark=o, mark options={solid, red}]
  table[row sep=crcr]{%
0.8	1\\
};

\end{axis}

\begin{axis}[%
width=0.326\fwidth,
height=0.494\fheight,
at={(0.337\fwidth,0\fheight)},
scale only axis,
axis on top,
xmin=-2.00869565217391,
xmax=2.00869565217391,
xtick={-2,-1,0,1,2},
xticklabels={\empty},
ymin=-2.00869565217391,
ymax=2.00869565217391,
ytick={-2,-1.5,-1,-0.5,0,0.5,1,1.5,2},
yticklabels={\empty},
axis background/.style={fill=white},
legend style={legend cell align=left, align=left, draw=white!15!black}
]
\addplot [forget plot] graphics [xmin=-2.00869565217391, xmax=2.00869565217391, ymin=-2.00869565217391, ymax=2.00869565217391] {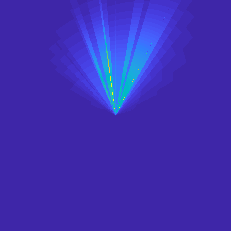};
\node[align=center, font=\color{white}]
at (axis cs:0,-1.4) {TTB obs.\\ \& unbiased model};
\addplot [color=red, draw=none, mark=o, mark options={solid, red}]
  table[row sep=crcr]{%
0.8	1\\
};

\end{axis}

\begin{axis}[%
width=0.326\fwidth,
height=0.494\fheight,
at={(0.674\fwidth,0\fheight)},
scale only axis,
axis on top,
xmin=-2.00869565217391,
xmax=2.00869565217391,
xtick={-2,-1,0,1,2},
xticklabels={\empty},
ymin=-2.00869565217391,
ymax=2.00869565217391,
ytick={-2,-1.5,-1,-0.5,0,0.5,1,1.5,2},
yticklabels={\empty},
axis background/.style={fill=white},
legend style={legend cell align=left, align=left, draw=white!15!black}
]
\addplot [forget plot] graphics [xmin=-2.00869565217391, xmax=2.00869565217391, ymin=-2.00869565217391, ymax=2.00869565217391] {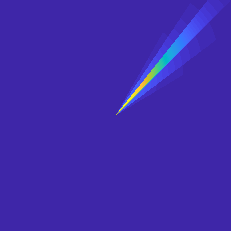};
\node[align=center, font=\color{white}]
at (axis cs:0,-1.4) {Unbiased obs.\\ \& unbiased model};
\addplot [color=red, draw=none, mark=o, mark options={solid, red}]
  table[row sep=crcr]{%
0.8	1\\
};

\end{axis}
\end{tikzpicture}%
\vspace{-10mm}\caption{Posterior densities $p(w \mid \mathcal{D}, \mathcal{H})$ computed over a grid of $(w_1, w_2)$ values and scaled such that the highest value in each is 1 (that is, the subfigures are not on the same scale). The true $w$ is indicated by a red circle.}
    \label{fig:lin_f_learning}
\end{figure}
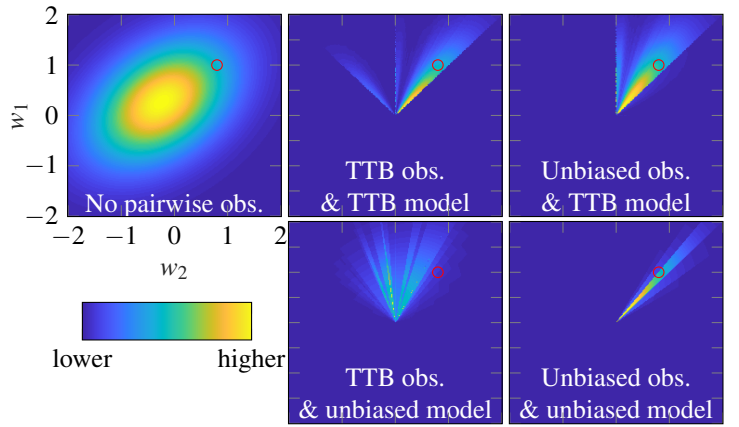

Figure~\ref{fig:lin_f_learning} shows the posterior densities for $w$ computed over a grid of points for a case with 2 training data points in $\mathcal{D}$ and $\mathcal{H}$ containing all pairwise comparisons of a randomly selected set of 10 points $x_i$ from a generated grid $X_G$, with true $w$ being $(1, 0.8)$. The TTB-generated pairwise observations help to concentrate the posterior density. However, if the biased nature of the observations is not accounted for, the main bulk of posterior mass misses the true value. For comparison, the figure also shows the result if we would have unbiased pairwise observations.

\section{Discussion}

We formulated a probabilistic, likelihood-based model of the Take The Best (TTB) heuristic. The decision strategy, that is, the cue search order, directions, and the decision uncertainty (or noise), arises from the posited model structure by conditioning on the data and using probabilistic inference (Bayes theorem). The learning performance of the model compared favourably to classic TTB, ABC-TTB, and logistic regression. This indicates, together with fast convergence of the MCMC computation to good configurations of cue order and directions, that the probabilistic TTB provides a computationally frugal formulation of the Take The Best decision strategy (although this particular computational strategy is not uniquely positioned; any that implements Bayesian inference would suffice). We also presented an extension to learning cue discrimination thresholds for continuous-valued cues. This moves the TTB heuristic towards a compensatory model, similar to the evidence accumulation model of \citet{lee2004evidence} that interpolates between non-compensatory (single cue) and compensatory (multi-cue) decision making by learning a stopping rule (evidence threshold). Indeed, the extension considerably increased the decision accuracy in the mileage dataset, the only dataset where the compensatory logistic regression outperformed the TTB models.

Our model states that cue order results from rational adaptation, where the agent learns the optimal ordering of cues, given the task environment (cues and choices).
This means that the principle of cue ordering is cognitively bounded, or computational, rationality.
Another principle that has been suggested is fluency, where the memory retrieval time or visual saliency determines the cue order \citep{dimov2017people}.
The principle of cognitively bounded rationality does not exclude the latter possibility.
In fact, its analysis of task behaviour that arises from the constraints of the environment \emph{and} the cognitive architecture readily accepts such constraints as memory or vision.
Above, we assumed perfect and immediate recall of cues, but it is entirely possible to extend the model to account for memory fluency.
The resulting agent would -- again, using rational adaptation -- adjust the cue order based on how readily they are available for recall.
Future work should add such constraints to the model, and investigate how the choice strategies change as their function.

We further demonstrated the benefit of formulating a probabilistic heuristic model by including it as a component in a function learning task to model preferential feedback from a biased agent. We believe that formulating joint probabilistic models of machine learning tasks and cognitive user models will be important, for example, in expert knowledge elicitation and interactive, human-in-the-loop machine learning \citep{daee2017knowledge, daee2018user}. Notably, the probabilistic formalism allows naturally capturing our evolving uncertainty about the user's decision strategy (e.g., cue search order) in interaction and propagate it to other parts of the system. This highlights a role for cognitive science in the next generation of machine learning systems that interact and learn together with human experts and end-users. Cognitive user models can be used to increase the performance of the systems and allow for more natural and efficient human--computer interaction. Important future work is to evaluate the presented approach with human decision data and interaction.

Source code for the probabilistic TTB and the experiments is available at \url{https://github.com/to-mi/pttb}.

\section{Acknowledgments}

This work was financially supported by the Academy of Finland (Finnish Center of Excellence in Computational Inference Research COIN, and Grants 295503, 294238, 292334, 284642, and 310947).

\bibliographystyle{apacite}

\setlength{\bibleftmargin}{.125in}
\setlength{\bibindent}{-\bibleftmargin}

\bibliography{references}

\end{document}